\documentclass[epj]{svjour}
\usepackage{epsfig,amsmath,amssymb,float}
\begin{document}
\newfloat{figure}{ht}{aux}

\title{Finite Size Scaling of the Spin stiffness of
the Antiferromagnetic $S=\frac{1}{2}$ XXZ chain}

\titlerunning{Finite Size Scaling of the Spin Stiffness of the XXZ chain}
\authorrunning{N. Laflorencie, S. Capponi and E. S. S\o rensen}

\author{Nicolas Laflorencie, Sylvain Capponi and Erik S. S\o rensen}

\institute{Laboratoire de Physique Quantique \& 
UMR CNRS 5626, Universit\'e Paul Sabatier, F-31062 Toulouse, France}  

\date{\today}

\abstract{We study the finite size scaling of the spin stiffness for the one-dimensional
$s=\frac{1}{2}$ quantum antiferromagnet as a function of the anisotropy
parameter $\Delta$. Previous Bethe ansatz results allow a 
determination of the stiffness in the thermodynamic limit. The Bethe 
ansatz equations for finite systems are solvable even in the presence of twisted boundary
conditions, a fact we exploit to determine the stiffness exactly for finite
systems allowing for a complete determination of the finite size corrections.
Relating the stiffness to thermodynamic quantities we calculate the 
temperature dependence of the susceptibility and its finite size corrections
at $T=0$.  A Luttinger liquid approach is used to study the finite size
corrections using renormalization group techniques and the results are
compared to the numerically exact results obtained using the Bethe
ansatz equations. Both irrelevant and 
marginally irrelevant cases are considered.}   

\maketitle


\section{Introduction}
The general XXZ Hamiltonian on a ring of size L is given by
\begin{equation}
\label{hamil}
H = H_0+H_i(\Delta)
 =
\frac{J}{2}\sum_{i=1}^L ( S_i^+S_{i+1}^- + hc ) + J\Delta\sum_{i=1}^L
 S_i^zS_{i+1}^z. 
\end{equation}
Here $H_0$ is the free part and $H_i(\Delta)$ is the interacting part. 
It is well known that this model is solvable when periodic boundary
conditions are applied~\cite{bethe31}. However, the same model can
also be solved under more general boundary conditions, in particular
under the so called twisted boundary conditions~\cite{hamer87,woyn87,alc88} defined by:
\begin{equation}
S^z_{L+1}=S^z_1,\ \  S^{\pm}_{L+1} = S^{\pm}_1 e^{\pm i\varphi},
\label{eq:htwist}
\end{equation}
where $\varphi$ is the twist angle. The application of such boundary 
conditions is equivalent to considering a system threaded by a magnetic
flux of strength $\varphi(\hbar c/e)$~\cite{byers61}. This fact was exploited by
Shastry and Sutherland~\cite{shas90,suth90} to calculate transport
properties of the system. Notably, the total ground state energy  as a
function of $\varphi$, $E_0(\varphi)$, can be calculated and hence also
the spin current and spin stiffness. In the thermodynamic limit they
showed that the spin stiffness is given by:
\begin{equation}
\label{eq:stifftherm}
\frac{\rho}{J}=\frac{\pi}{4}\frac{\sin(\mu)}{\mu (\pi - \mu)},\ \Delta=\cos(\mu).
\end{equation}
Hence, for the antiferromagnetic Heisenberg model, $\Delta=1$,
$\rho/J=1/4$. In the context of mesoscopic physics it is quite interesting
to study transport properties for finite systems where coherence effects 
are important~\cite{ian01} and notably the finite size dependence of the
current, susceptibility and stiffness are important for a complete
understanding of the experimental results. The stiffness, $\rho$, of the system is
closely related to the dc conductivity of the system and a non-zero
$\rho$ implies quasi long-ranged correlations with power-law decay.  
In the present paper we show that it is possible to calculate numerically exactly 
the spin stiffness, $\rho(L)$, for a finite system. $\rho(L)$ can then
be used to calculate the susceptibility for finite systems and finite
temperatures. Using a Luttinger liquid approach it is possible to
understand quite completely the structure of the finite-size corrections
and we compare these perturbative results to the numerical ones.

We consider exclusively the antiferromagnetic (AF) case with $J=1$ and
we shall mainly be concerned with the regime where the anisotropy parameter,
$\Delta$, lies between $\Delta=0$ (XY) and $\Delta=1$ (XXX). It is well known
that in the regime $\Delta\in [-1,1]$ this model display gapless excitations
with power law correlations and off-diagonal long-range order characterized 
by a non-zero stiffness, $\rho$, in close analogy to the superfluid order
parameter in the two-dimensional {\it classical} $XY$ model. For $\Delta > 1$
the model Eq.~(\ref{hamil}) enters a phase with Ising like AF order and a
non-zero gap. Following the above remarks, this transition, occurring at 
$\Delta=1$, can be viewed as a metal-insulator transition~\cite{shas90}.
We concentrate on the region $\Delta\in[0,1]$ with emphasis on the
finite-size corrections at $\Delta=1$. We first briefly review the
finite-size scaling of the stiffness in section~\ref{sec:stiff}, then 
we discuss the simple free case, $\Delta=0$, (XY). Section~\ref{sec:rg}
presents the renormalization group (RG) approach leading to the predictions
for the finite-size scaling behavior discussed in
section~\ref{sec:fscaling}
considering both irrelevant ($\Delta\in\lbrack0,1\lbrack $ ) and
marginally irrelevant
($\Delta=1$) cases. Section~\ref{sec:ba}-\ref{sec:numres} present the
numerically exact method for calculating the stiffness (and implicitly
the susceptibility) and the numerical
results up to $L=10000$ are compared to the RG predictions from the previous
sections.

\section{Scaling of Stiffness}\label{sec:stiff}
Suppose a twist of size $\varphi$ is applied at the boundary of 
an otherwise uniform system. It is natural to expect that this will give
rise to a uniform phase gradient
\begin{equation}
\nabla\theta=\varphi/(aL),
\end{equation}
throughout the system with $L$ sites and lattice spacing $a$. We can now define the stiffness
with respect to the resulting change in the ground-state energy density in the
following way~:
\begin{equation}
\frac{\delta e_0}{\hbar}=\frac{1}{2}\rho(\nabla\theta)^2,
\end{equation}
where $\delta e_0$ is the change in the ground-state energy density
when the twist $\varphi$ is applied. It then follows that the stiffness
is given by the following expression~:
\begin{equation}
\rho(L)=\frac{(aL)^2}{\hbar}
\frac{\partial^2e_0(\varphi)}{\partial\varphi^2}|_{\varphi=0}.
\label{eq:stiffness}
\end{equation}
Hence, $\rho$ has dimension of inverse (length)$^{d-2}\times$t.
In the vicinity of a quantum critical point or line we can invoke hyperscaling
and two-scale factor universality (see reference~\cite{wallin} for a
discussion and references) to argue that
\begin{equation}
\rho\xi^{d-2}\xi_\tau=C,
\end{equation}
where $C$ is a {\it universal} constant. Applying standard finite size scaling
theory~\cite{privman} we then expect the stiffness to obey the following finite
size scaling ansatz~\cite{wallin}:
\begin{equation}
\rho(L)=L^{-d-z+2}\tilde\rho(L^{1/\nu}\delta),
\label{eq:fscaling}
\end{equation}
where $\delta$ is the distance to the quantum critical point. In a phase with
long-range order, the stiffness should diverge and in the absence of long-range
order we expect $\rho$ to vanish exponentially with the system size. Hence, 
Eq. (\ref{eq:fscaling}) describes the finite-size corrections close to 
a critical point.
In the present
case of the one-dimensional Heisenberg chain we expect to have $z=1$
and consequently $-d-z+2=0$. Since the twist is applied in the $XY$ 
component of the spins we expect the resulting stiffness to be non-zero
and universal
in the critical region between $\Delta\in[-1,1]$. Eq.~(\ref{eq:fscaling})
then tells us that the leading finite-size corrections in this region are
absent and in the thermodynamic limit we expect $\rho$ to jump
discontinuously at $\Delta=-1,1$, as noted in Ref.~\cite{shas90}.
It is important to note that the above finite-size scaling analysis 
do not take into account corrections to scaling which, as we shall
see later on, are especially important close to $\Delta=1$.

\section{Free case, $\Delta=0$}\label{sec:free}

At $\Delta=0$, $H=H_0$ and the Hamiltonian (\ref{hamil}) becomes equivalent to a system
of free fermions that can be diagonalized explicitly in
k-space through the use of the Jordan-Wigner
transformation  $S^z_j = 1/2-n_j$, and $S^\dagger_j =\psi_j
e^{i\pi\sum_{l=1}^{j-1}n_l}$. The $\psi_j$ satisfies fermionic commutation
relations, $ \lbrace\psi^\dagger_i,\psi_j\rbrace
=\delta_{ij}$, and  $n_j = \psi^\dagger_j\psi_j$ is the number of fermions
(spin down) at the j-site.  We can now write
$H_0$ in k-space
\begin{eqnarray}
\label{h0}
H_0=-J \sum_k\cos(k)\psi^\dagger_k\psi_k. 
\end{eqnarray}
The antiferromagnetic ground state is in the $S^z_{tot} =0$ sector, which
corresponds to half-filling ($n_{tot} =L/2$). The effect of the twist
$\varphi$ is a shift in the momentum space: $k\longrightarrow k+\varphi$, this can
also be seen explicitly from the Bethe ansatz equation (see
section~\ref{sec:ba}). 
Therefore, we obtain a rather simple expression for the ground state energy per site 
versus $\varphi$~:  
\begin{eqnarray} 
\frac{e_0 (L,\varphi)}{J}=-\frac{1}{L}\frac{\cos(\frac{\varphi}{L})}{\sin(\frac{\pi}{L})}.
\label{e0} 
\end{eqnarray} 
Using Eq.~(\ref{eq:stiffness}), the finite size corrections for the stiffness
are then~:
\begin{eqnarray} 
\label{r0L} \frac{\hbar}{J a^2}\rho (L) =\frac{1}{L\sin(\pi/L)}\simeq\frac{1}{\pi}+\frac{\pi}{6L^2}+
{\cal O}(\frac{1}{L^4}).
\label{eq:rhoLfree}
\end{eqnarray} 
The corrections of order $o(\frac{1}{L^2})$ can be
understood in terms of conformal field theory~\cite{cardy84}. We also have  
the universal corrections to the free energy which are given by
$e_0(L)=e_0^{\infty}-\frac{{\pi}uc}{6L^2}+{\cal O}(\frac{1}{L^4})$. Here, $u$ is the velocity of
excitations ($u=1$ in the free case) and $c$ is the conformal anomaly number;
$c=1$ for the $s=1/2$ XXZ chain~\cite{affleck89}.

\section{Renormalization group analysis}\label{sec:rg}
\subsection{Bosonization of the XXZ chain}
In this section, we introduce the bosonization of interacting fermionic
systems like the XXZ chain (using Jordan-Wigner transformation) or the
Hubbard model, which are Luttinger liquids \cite{Schulz98,voit95}. In a
few words, it consists in linearizing the dispersion relation near the
Fermi points and studying the low energy excitations close to those
points. Moreover, for commensurate fillings of order $n$, umklapp
scattering ($n$ left electrons become right 
or vice-versa) is possible and gives rise to an additional Sine-Gordon term in the
Hamiltonian. In the continuum limit and in the $S^z_{tot} =0$ sector
(corresponding to half-filling), the XXZ chain is governed by a 
Hamiltonian expressed in term of conjugate boson fields $\Phi(x)$ and $\Pi(x)$ as
follows~:
\begin{eqnarray}
\label{sinegor}
H &=&\int \frac{dx}{2}\lbrace(\frac{uK}{\pi})\lbrack{\pi\Pi(x)}\rbrack^2+
(\frac{\pi u}{K})\lbrack\frac{\partial_x \Phi(x)}{\pi}\rbrack^2\rbrace
\nonumber \\
&+& \frac{J\Delta}{2(\pi a)^2}\int{dx\cos(4\Phi(x))}.    
\end{eqnarray}
The last term represents umklapp scattering ans $u$, $K$ are the two
Luttinger liquid parameters which describe the low-energy physics.  

If the ring is threaded by a flux $\varphi$, the effect is just a shift
of the current $\Pi(x)$ like $\Pi(x)\longrightarrow
\Pi(x)+\frac{\varphi}{L\pi}$~\cite{Schulz98}. Since we are dealing here with fermions
carrying a charge, we can identify the factors entering the Hamiltonian as
the charge stiffness, given by $\frac{uK}{\pi}$, and the compressibility $\frac{K}{u\pi}$. Note that we are working now in units where $\hbar=1,
a=1$.
 
Using Jordan-Wigner transformation, we obtain, in the spin chain case the
spin stiffness, 
\begin{equation}
\rho =\frac{uK}{\pi}, 
\end{equation}
and the susceptibility, 
\begin{equation}
\chi =\frac{K}{u\pi}.
\end{equation}
The susceptibility can then be obtained from $u$ and $\rho$ using the 
hydrodynamic relation~\cite{shas90}:
\begin{equation}
\chi = \frac{\rho}{u^2}. 
\label{eq:hydro}
\end{equation}
In the thermodynamic limit, we will note $u^*$ and $K^*$ for the
Luttinger liquid parameters. By using the thermodynamic relations from above
and comparing to exact results obtained from Bethe ansatz, we can express 
these parameters versus $\Delta=\cos(\mu)$~:
\begin{eqnarray}
\frac{u^*(\Delta)}{J}=\frac{\pi}{2} \frac{\sin(\mu)}{\mu},\nonumber\\
K^*(\Delta)=\frac{\pi}{2(\pi-\mu)}.
\end{eqnarray}

\subsection{RG equations}
\begin{figure}
\begin{center}
\epsfig{file=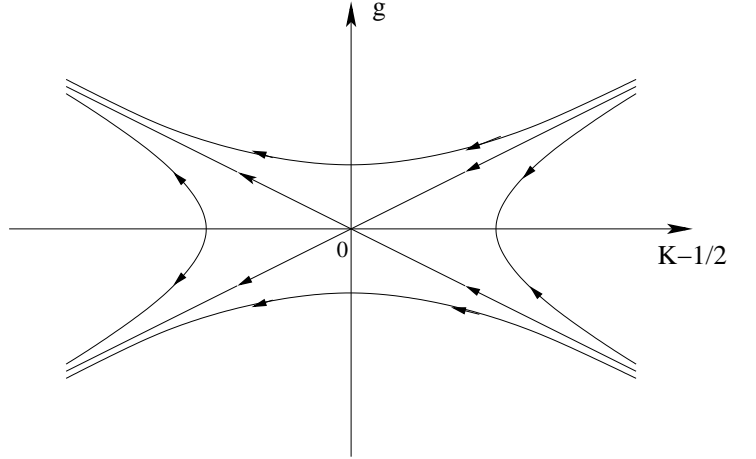,width=10cm}
\caption{RG flow of the Sine-Gordon model.}
\label{fig:KTrg}
\end{center}
\end{figure}

Now, we have to study the Sine-Gordon term in Eq.~(\ref{sinegor}) ($g\int
{dx\cos(4\Phi(x))}$) as a perturbation and examine the effect of a renormalization under a
change of length scale. We then obtain~: 
\begin{eqnarray}
\label{KT}
\frac{dg}{dl} & = & (2-4K)g +{\cal O}(g^2) \\
\frac{dK}{dl} & = & -Ag^2,   \nonumber 
\end{eqnarray}
where $g=J\Delta/(2(\pi a)^2)$ is the coupling, $l=\ln(L)$ and A is a constant.  Those
equations are identical to the Kosterlitz-Thouless renormalization group
analysis~\cite{KT73} used in the classical 2d XY model and  which gives a
description of the superfluid transition. There is a powerful analogy
between the superfluid density which vanishes at the transition and the
spin stiffness in the quantum XXZ chain. 
Here, the renormalization is done versus the chain length $L$ at $T=0$,
but due to the invariance of the model it should be equivalent to
considering an infinite chain at $T= u/L$
\cite{Affleck94}. 
The renormalization flow diagram for this model is shown in Fig.~\ref{fig:KTrg}
where distinct behaviors are observed~:   
\begin{itemize}
\item For $1/2<K\leq1$ ($\Delta\in\lbrack0,1\lbrack $ ), the
perturbation $g$ remains irrelevant and vanishes rapidly with the size L, and
$K$ goes to $K^*$. 
\item At the isotropic point $K=1/2$ ($\Delta=1$), the
perturbation is marginally irrelevant implying a  
logarithmically decreasing $g$ and $K\to K^*$.
\item For $K<1/2$ ($\Delta >1$), the umklapp term, $g$,  increases with
the system size because the perturbation is relevant and $K$ vanishes. A
 gap opens up in the spectrum and the stiffness falls to zero.
\end{itemize}

\section{Finite size scaling}\label{sec:fscaling}
The integration of the RG equations~(\ref{KT}) from $L_0$ to $L$ gives us the finite-size scaling of $K(L)$ and $g(L)$. The corrections to $g$ have already
been evaluated by Cardy~\cite{cardy86} and by Lukyanov~\cite{Lukya98}. 
They are very important to
calculate the corrections to the energy \cite{affleck89,Lukya98,Nomura93}. At the
isotropic point, the coupling $g$ is marginally irrelevant and the
integration of Eq.~(\ref{KT}) up to ${\cal O}(g^2)$ gives rise to logarithmic
corrections to the ground-state energy~\cite{Nomura93}~:
\begin{equation}
\label{e(L)}
e_0 (L) = e_0^{\infty}-\frac{\pi u}{6L^2}[c+8\pi^3 bg(L)^3],
\end{equation}
where $c=1$ and $b=4/\sqrt{3}$ is determined by a three-point function\cite{affleck89}.
Here, the corrections of $K$ will imply corrections to the stiffness and susceptibility
in two cases: when the perturbation is irrelevant and
marginally irrelevant.

\subsection{Irrelevant perturbations: power-law corrections}
In the anisotropic antiferromagnetic regime ($1/2<K^*\leq1$), we can integrate
Eq.~(\ref{KT}) from $L_0$ to $L$, assuming the
integration with $L\gg L_0$. We then obtain the finite size corrections to $K$~:
\begin{eqnarray}
\label{irrK}
K(L)&=&K^*+ \frac{(K(L_0)-K^*)(1-2K^*)}{1-K^* -K(L_0)}
\left(\frac{L}{L_0}\right)^{-8(K^*-1/2)} \nonumber\\
&+&{\cal O}\left(\frac{L}{L_0}\right)^{-16 (K^*-1/2)}.
\end{eqnarray}
We see that the exponent of 1/L lies between $0^+$ and 4. In the free case
($K=1$), we find an exponent 4 which seems to contradict the finite-size
corrections found in Eq.~(\ref{r0L}). However, it must be noted that terms in
$1/L^2$, the so-called 'analytical' corrections, are expected to appear in
all quantities since they are related to the conformal symmetry of the
fixed-point Hamiltonian~\cite{cardy84,voit95}. 
We can also emphasize that these analytical corrections  dominate when 
 $K\geq 3/4$ (i.e $\Delta\leq 1/2$) and are sub-dominant
 otherwise. This will be shown precisely with the 
numerical Bethe ansatz calculation. 

{F}rom Eq.~(\ref{irrK}), we expect that to the lowest order, 
the finite size corrections  for the stiffness and the susceptibility are~:
\begin{align}
\label{irrrho}
{\mathrm{If}\ } 1/2<\Delta <1&:&\nonumber\\
\rho(L)-\frac{uK^*}{\pi} &\sim& \chi(L)-\frac{K^*}{u\pi} \sim
\left(\frac{1}{L}\right)^{8(K^*-1/2)},\nonumber\\
{\mathrm{If}\ } 0\leq\Delta \leq1/2&:& \nonumber\\
\rho(L)-\frac{uK^*}{\pi} &\sim&  \chi(L)-\frac{K^*}{u\pi} \sim
\left(\frac{1}{L}\right)^2.\qquad\qquad
\end{align}

\subsection{Marginal irrelevant perturbations: logarithmic corrections}
At the isotropic point, we can integrate  Eq.~(\ref{KT}) at the lowest
order, taking  initial
conditions at $L_0$. We find that $K$ is decreasing logarithmically~:
\begin{equation}
\label{margK}
K(L)=\frac{1}{2}+\frac{K(L_0)
-\frac{1}{2}}{1+4(K(L_0)-\frac{1}{2})\ln(\frac{L}{L_0})}.
\end{equation}
Such behavior is in agreement with the logarithmic corrections obtained by Loss {\it et al.}~\cite{Loss94}. 
At first order in $1/\ln(L/L_0)$, the finite size scaling correction for the stiffness
and the susceptibility are
\begin{equation}
\label{margrho}
\frac{\rho(L)}{J}\simeq \frac{1}{4}+\frac{\rho(L_0)/J-1/4}{1+8(\rho(L_0)/J-1/4)\ln(L/L_0)},
\end{equation}
\begin{equation}
\label{margchi}
J\chi(L)\simeq
\frac{1}{\pi^2}+\frac{J\chi(L_0)-1/\pi^2}{1+2\pi^2(J\chi(L_0)-1/\pi^2)\ln(L/L_0)}. 
\end{equation}
As noted previously, these results also give the temperature dependence of
these quantities by taking $T=u/L$ and we obtain results in agreement with
previous work by Eggert {\it et al.}~\cite{Affleck94}. 
These predictions will be compared to numerically exact results in section~\ref{sec:numres}.

\section{Bethe Ansatz Equations}\label{sec:ba}
The solution of the one-dimensional antiferromagnetic Heisenberg
model due to Bethe~\cite{bethe31} can be extended to quite general
boundary conditions. Hamer, Quispel and Batchelor~\cite{hamer87} exploited
this fact to calculate the ground state energy in the thermodynamic
limit for the Hamiltonian with
twisted boundary conditions, Eq. (\ref{eq:htwist}), as a function of the
applied twist $\varphi$.  Using this result
the stiffness of the system can be calculated~\cite{shas90}
using the relation Eq.~(\ref{eq:stiffness}).
If twisted boundary conditions are imposed the expression for the 
ground state energy per site
remains unchanged and are the same as for the uniform case:
\begin{equation}
\frac{e_0(L,\varphi)}{J}=\frac{\Delta}{4}-\frac{1}{L}\sum_{l=1}^{L/2}\cos(k_l (\varphi)).
\label{eq:Bags}
\end{equation}
The change in the boundary conditions enters in the equations
determining the $L/2$ quasi-momenta, $k_l$, only in the following manner:
\begin{equation}
k_l = \frac{1}{L}\lbrack 2\pi I_l + \varphi - \sum_{n\neq l}\Theta_{l,n}\rbrack,
\label{eq:kl}
\end{equation}
where $\Theta_{l,n}$ is given by:
\begin{equation}
\Theta_{l,n} = 
2\arctan\lbrack\frac{\Delta\sin(\frac{k_l -k_n}{2})}
{\cos(\frac{k_l+k_n}{2})-\Delta\cos(\frac{k_l-k_n}{2})}\rbrack.
\end{equation}
For the ground state, the set of integers $I_l$ is given by 
$I_l=l-(L/2+1)/2,\ \ l\in\lbrack 1,L/2\rbrack$.
Using the above equations it is possible to numerically determine the
quasi-momenta, $k_l$, even for relatively large systems and for non-zero
$\varphi$.

Using Eq.~(\ref{eq:stiffness}) and Eq.~(\ref{eq:Bags} ), 
we can now formally write an equation for $\rho(L)$:
\begin{equation}
\frac{\rho(L)}{J}=L\sum_{l=1}^{L/2}
\lbrack\frac{\partial^2 k_l}{\partial\varphi^2}\sin(k_l)+
(\frac{\partial k_l}{\partial\varphi})^2
\cos(k_l)\rbrack|_{\varphi=0}.
\label{eq:rhoL}
\end{equation}
Hence it is possible to rather easily calculate $\rho(L)$ if the
derivatives ${\partial^2 k_l}/{\partial\varphi^2}$ and
${\partial k_l}/{\partial\varphi}$ can be calculated. For
${\partial k_l}/{\partial\varphi}$ we obtain the following expression:
\begin{equation}
\frac{\partial k_l}{\partial\varphi}=
\frac{1}{L}-\frac{1}{L}\sum_{n=1}^{L/2}\lbrack
\frac{\partial\Theta(k_l,k_n)}{\partial k_l}\frac{\partial
k_l}{\partial\varphi}+
\frac{\partial\Theta(k_l,k_n)}{\partial k_n}\frac{\partial
k_n}{\partial\varphi}\rbrack.
\end{equation}
Since the derivatives ${\partial\Theta(k_l,k_n)}/{\partial k_l}$ do not
depend on ${\partial k_l}/{\partial\varphi}$, but only on the previously
determined quasi-momenta $k_l$, this is a simple matrix equation from 
which ${\partial k_l}/{\partial\varphi}$ can be determined using
standard linear algebra routines. An equivalent expression exists for
${\partial^2 k_l}/{\partial\varphi^2}$ which also reduces to a linear
algebra problem once ${\partial k_l}/{\partial\varphi}$ is known. Hence,
$\rho(L)$ can be determined numerically exactly once the $k_l$'s have 
been obtained.

\section{Numerical results}\label{sec:numres}
\begin{figure}
\begin{center}
\epsfig{file=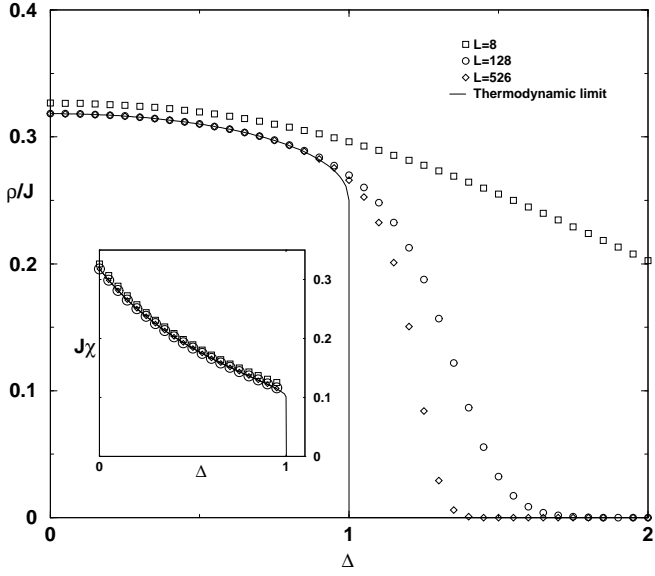, width=\columnwidth}
\caption{The spin stiffness $\rho(L)$ of the antiferromagnetic 
chain as a function of the anisotropy $\Delta$ for different ring sizes, $L$. 
The solid line is the result Eq.~(\ref{eq:stifftherm}). In the
inset is shown the uniform susceptibility versus $\Delta$ obtained from
Eq.~(\ref{eq:hydro})}.
\label{fig:stiff}
\end{center}
\end{figure}
We begin by a discussion of our results for the spin stiffness,
$\rho(L)$, for finite rings of size $L$.
In Fig.~\ref{fig:stiff} our results for the spin stiffness are shown as
a function of the anisotropy $\Delta$ for different ring sizes, $L$.    
As noted in section~\ref{sec:stiff}, we expect the leading finite size
correction to be absent in the gapless regime $\Delta\in\lbrack
0,1\lbrack$. This is rather clearly the case. Only when $\Delta>1$ do 
these corrections become important, the system opens up
a gap and the correlation length, $\xi$, becomes finite.
In this regime, an explicit expression for $\xi$ has been obtained by
Baxter~\cite{baxter82} and due to this finite correlation length we 
expect $\rho(L)$ to vanish in an exponential manner with the system
size, $\rho(L)\sim e^{-L/\xi}$.
For $\Delta< 1$, the
finite size corrections are expected to have the form (\ref{irrrho}). Finally, at $\Delta=1$, we expect logarithmic corrections of
the form (\ref{margrho}). These logarithmic corrections are
non-negligible and remain sizable for systems of macroscopic size.
Hence, the discontinuous jump in $\rho$ at $\Delta=1$ is difficult to
observe.

\subsection{Power-law corrections}
We now turn to a discussion of our results for the finite size
corrections in the critical region $\Delta<1$. Following our results
from the previous sections, we expect $\rho(L)$ to be independent of
$L$ to leading order and the corrections to scaling to acquire a
power-law dependence with an exponent depending on $K^*$.

\begin{figure}
\begin{center}
\epsfig{file=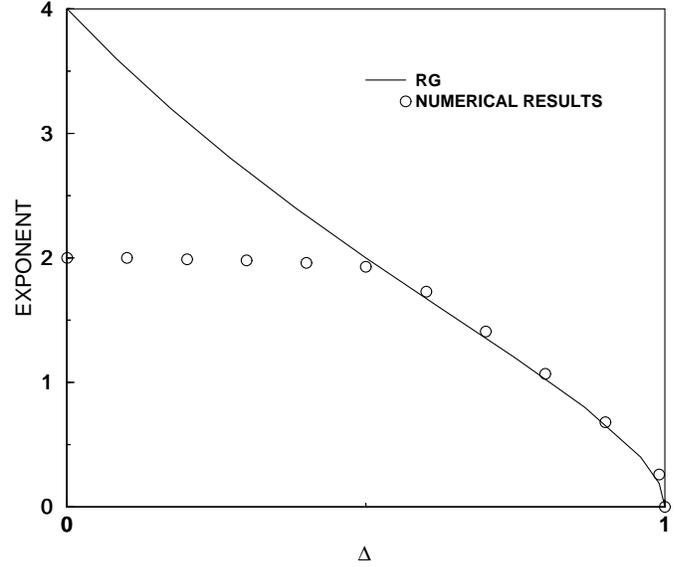,width=\columnwidth}
\caption{The numerically determined exponent of the leading correction 
to scaling in $1/L$ 
of $\rho(L)$ as a function of the anisotropy $\Delta$ (circles). 
The solid line indicates the RG result for this exponent 
$8(K^*-1/2)=\frac{4\arccos(\Delta)}{\pi-\arccos(\Delta)}$.
}
\label{fig:expos}
\end{center}
\end{figure}
Writing $\rho(L)=\rho+{\cal O}(L^{-\alpha})$, with $\rho$ given by
Eq.~(\ref{eq:stifftherm}), we can determine the exponent $\alpha$ from
the numerically exact results obtained from the Bethe ansatz. We have
determined this exponent in the region $\Delta<1$, and our results are
shown in Fig.~\ref{fig:expos}. The solid line indicates the
renormalization group result, Eq.~(\ref{irrK}).
As previously explained we expect on general grounds always to have
corrections to scaling of the form $1/L^{2}$, as explicitly shown in
Eq.~(\ref{eq:rhoLfree}). Corrections of such a form dominate when
$0<\Delta<1/2$. Accordingly, the corrections to scaling coming from the
marginally irrelevant coupling are dominant when $1/2<\Delta<1$.

\subsection{Logarithmic corrections}
\begin{figure}
\begin{center}
\epsfig{file=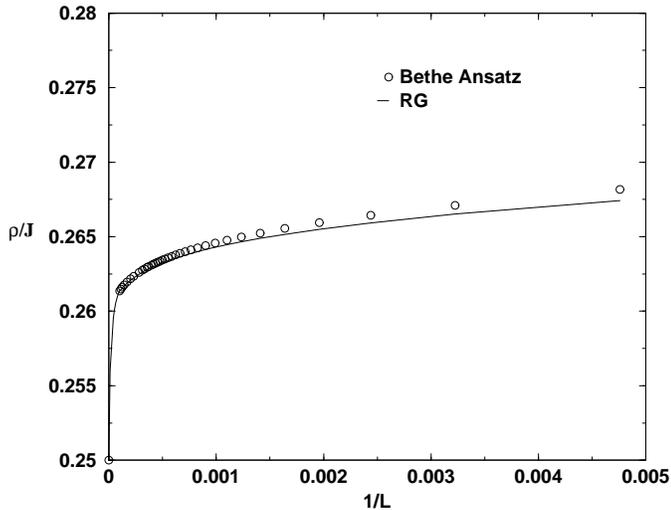,width=\columnwidth}
\caption{Spin stiffness, $\rho(L)$ as a function of 1/L at $\Delta=1$. 
The RG result Eq.(\ref{margrho}) (solid line), with $L_0 =10000$ is compared 
to exact numerical results from the Bethe ansatz ($\circ$).
The thermodynamic limit value, $\rho(1/L=0)/J=1/4$, has been obtained from Eq.
(\ref{eq:stifftherm}). }
\label{fig:rhoLHAF}
\end{center}
\end{figure}

Finally we discuss our results for the isotropic point.
At $\Delta=1$, the perturbation to the fixed point Hamiltonian is marginally
irrelevant, giving rise to finite size corrections of logarithmic
form for the spin stiffness, Eq.~(\ref{margrho}), and 
the susceptibility, Eq.~(\ref{margchi}). 
In order to compare our numerical results to the RG results we use our
results for the largest possible system size as the reference point,
$L_0$, and perform the comparison for $L<L_0$
Taking $L_0=10000$, we show in Fig.~\ref{fig:rhoLHAF} our results for 
stiffness as a function of system size, $L$. The solid line indicates
the RG result for the stiffness, Eq.~(\ref{margrho}), and 
the numerical results from the Bethe ansatz solution are shown as circles.
Numerical calculations have been performed for system sizes up to 10000 sites.
A good agreement between the RG results and the numerical results is
evident even down to rather small system sizes. From
Eq.~(\ref{eq:hydro}) we see that the susceptibility follows a similar
form. 

\begin{figure}
\begin{center}
\epsfig{file=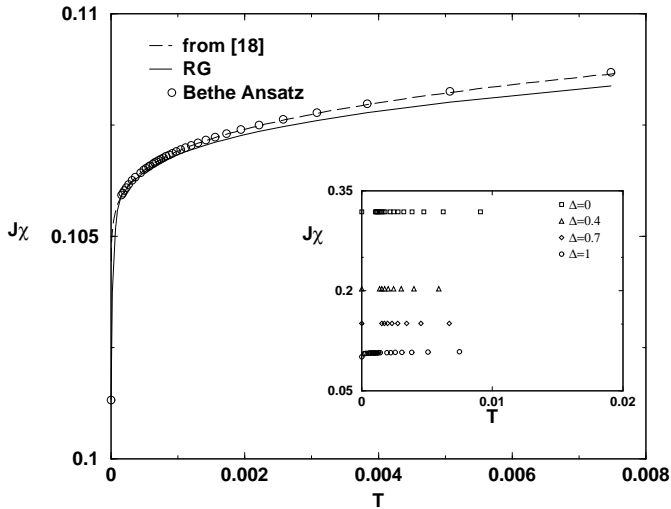,width=\columnwidth}
\caption{Susceptibility $\chi(T)$ as a function of the temperature $T$. The 
circles indicate the exact Bethe ansatz results, the solid line is the RG result
Eq.~(\ref{eq:suscT}) with $T_0=\pi/(2*L_0)$ and the dashed line the result from
Ref.~\cite{Lukya98}. The inset shows the temperature
dependence of the susceptibility for different values of the anisotropy $\Delta$.
For $\Delta\neq 1$, the susceptibility follows a power law form with an
exponent of 2 at $\Delta=0$ and $0.4$.  At $\Delta=0.7$ the exponent is close
to $8(K^*-1/2)\simeq 1.356$ as expected.}
\label{fig:susc}
\end{center}
\end{figure}
Using the relationship $L\longleftrightarrow u/T$, we can obtain
information about the temperature dependence of the susceptibility
in the thermodynamic limit from our results obtained
at $T=0$ for finite systems.
The temperature dependence of the Drude weight has recently been
calculated using
the thermodynamic Bethe ansatz~\cite{zotos} and is currently a topic of
discussion~\cite{alvarez}.
Previously, the low temperature behavior of the susceptibility has also been 
studied in detail by Eggert et al.~\cite{Affleck94} and
Lukyanov~\cite{Lukya98}.
In analogy with the work of Eggert et al.~\cite{Affleck94}, we obtain from Eq.(\ref{margchi})
the following expression for the low temperature dependence of the susceptibility 
in the thermodynamic limit~:
\begin{equation}
J\chi(T)\simeq \frac{1}{\pi^2}+
\frac{J\chi(T_0)-1/\pi^2}{1+2\pi^2(J\chi(T_0)-1/\pi^2)\ln(T_0/T)}.
\label{eq:suscT}
\end{equation}
As previously done, we use our largest system size to define an
equivalent temperature, $T_0$, and use the RG expression
Eq.~(\ref{eq:suscT}) for higher temperatures (smaller system sizes).
Taking $T_0 =u/L_0=\pi J/(2*L_0)\sim 1.57\times10^{-4}$ ($L_0=10000$), 
we show in Fig.~\ref{fig:susc} the RG result, Eq.~(\ref{eq:suscT}) (solid
line) along with the exact numerical results from the Bethe ansatz
(circles) and the result of Ref.~\cite{Lukya98}.
As already shown by Eggert et al.~\cite{Affleck94} for the isotropic
case, the agreement between the RG result and the Bethe ansatz results
is excellent at low temperatures (large system sizes). However, the
result of Lukyanov~\cite{Lukya98} seems to work better over the complete
temperature range. The logarithmic
dependence of the susceptibility at $\Delta=1$ is clearly visible and
the derivative with respect to temperature of $\chi(T)$ diverges as
$T\to 0$ and $\chi(T)\to 1/\pi^2$. 

As pointed out in the previous
sections, we expect a power-law dependence of $\chi(T)$ with the temperature 
for $\Delta<1$ with an exponent of 2 for $\Delta<1/2$ and a non-trivial 
exponent in the regime $1/2<\Delta<1$. In the inset of
Fig.~\ref{fig:susc}, $\chi(T)$ is shown for several different
values of $\Delta<1$. For $\Delta>1$ the system opens up a gap and the
susceptibility decreases exponentially at low temperatures.

\subsection{Coupling term $g$}
\begin{figure}
\begin{center}
\epsfig{file=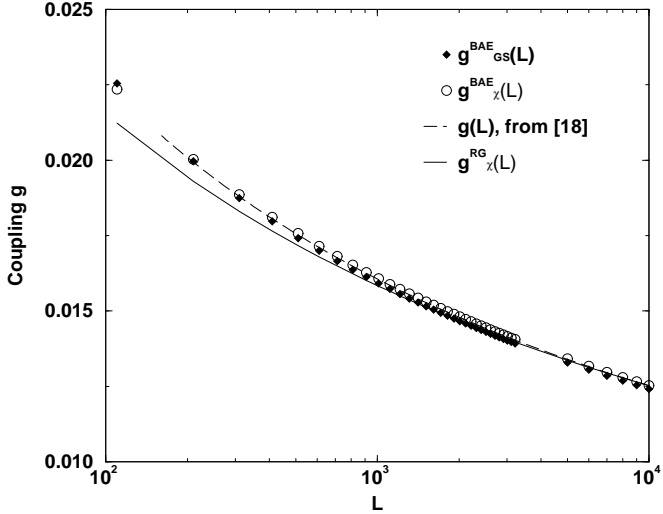,width=\columnwidth}
\caption{The flow of the coupling constant $g(L)$ with system size $L$
for the isotropic Heisenberg model.
The solid line indicates the RG result, Eq.~(\ref{eq:gRG}). The symbols
indicate estimates of $g(L)$ from the scaling of the ground state
energy and susceptibility and the dashed line represents the result
Eq.~\ref{eq:glukya}.}
\label{fig:gL}
\end{center}
\end{figure}

{F}rom our results for the susceptibility (or equivalently the stiffness)
it is possible to obtain an estimate of the flow of the
Sine-Gordon coupling term $g(L)$ with the system size $L$. 
Previous studies have obtained the same flow from the 
ground state energy~\cite{affleck89}.
At the isotropic point ($\Delta=1$), we have seen
that the perturbation term in Eq.(\ref{sinegor}), $g\int {dx\cos(4\Phi(x))}$, is
marginally irrelevant. Hence, we expect a logarithmic behavior of $g(L)$.
From the results of Eggert et al.~\cite{Affleck94} for the susceptibility in
the $k=1$ WZW non-linear $\sigma$ model~\cite{Affleck90}, the coupling $g$ can be expressed in term of
$\chi(L)$ obtained numerically from the solution of the Bethe ansatz
equations as follows:
\begin{equation}
g^{BAE}_{\chi}(L)=\frac{\pi\sqrt3}{2}(J\chi(L) - \frac{1}{\pi^2}).
\label{eq:gchiL}
\end{equation}
As mentioned, it is also possible to estimate $g(L)$ from the finite
size corrections to the ground state energy~\cite{affleck89}, Eq.~(\ref{e(L)}).
In this case one finds:
\begin{equation}
g_{GS}^{BAE}(L)=\left[\frac{12L^2/\pi^2(e_0^{\infty}-
e_0(L))-1}{32\pi^3/\sqrt3}\right]^{1/3},
\label{eq:gGS}
\end{equation} 
where $e_0(L)$ again is determined from the numerical solution of the
Bethe ansatz equations.
Finally, we can compare these two estimates to the RG result for 
$g(L)$ which is given by the same expression than Eq.~(\ref{eq:gchiL})
except that $\chi(L)$ is given by the RG expression Eq.~(\ref{margchi})~: 
\begin{equation}
g_{\chi}^{RG}(L)=\frac{\pi\sqrt3}{2}\frac{J\chi(L_0)-
1/\pi^2}{1+2\pi^2(J\chi(L_0)-1/\pi^2)\ln(L/L_0)}.
\label{eq:gRG}
\end{equation} 
In this equation, we use $\chi(L_0)$ to determine the bare coupling
$g_0$.
We can also compare our results to the
general expression, including higher order corrections, derived by 
Lukyanov \cite{Lukya98} for $g(L)$ which is, in our notation:
\begin{equation}
g^{-1}+\frac{\sqrt3}{8\pi} 
\ln(g)=\frac{\sqrt3}{8\pi} 
\ln(\frac{2\sqrt2}{3^{0.25}}e^{\gamma+0.25}\times L),
\label{eq:glukya}
\end{equation}
where $\gamma$ is the Euler constant.
In Fig.~\ref{fig:gL} we show results for $g_{GS}^{BAE}(L)$ 
(solid diamonds) and $g_{\chi}^{BAE}(L)$ ($\circ$).
The solid line indicates the RG result, Eq.~(\ref{eq:gRG}) and the
dashed line indicates the result of Lukyanov~\cite{Lukya98}.
The largest system size, $L_0=10000$, has again been used to determine
the bare parameters in Fig.~\ref{fig:gL}.
As evident from the results shown in Fig.~\ref{fig:gL} there is an
excellent agreement between the numerical estimates for $g(L)$ obtained
from the ground state energy $e_0(L)$ and the susceptibility, $\chi(L)$
and the RG result, Eq.~(\ref{eq:gRG}).
The result of Lukyanov~\cite{Lukya98} agrees very well with the
numerical results over the entire range of $L$.

\section{Conclusion}
The finite size corrections to the spin stiffness and subsequently the
susceptibility have been studied in detail for the $S=1/2$
antiferromagnetic spin chain as a function of the anisotropy parameter
$\Delta$. At the isotropic point, $\Delta=1$ where logarithmic
corrections are dominant, our results are in good agreement with
previously known results. From the finite size dependence of the
spin stiffness and susceptibility we are able to obtain precise results
for the low temperature behavior of the susceptibility in the
thermodynamic limit. The described effects should be observable in 
mesoscopic systems where a detailed understanding of the finite size
behavior is crucial.

We thank P.~Azaria for useful discussions.


\begin{thebibliography}{}

\bibitem{bethe31} H.~Bethe , Z. Physik {\bf 38}, 441 (1931). 

\bibitem{hamer87} C.~J.~Hamer, G.~R.~W.~Quispel, and M.~T.~Batchelor, 
J.~Phys.~A {\bf 20}, 5677 (1987).

\bibitem{woyn87} F.~Woynarovich and H.-P.~Eckle,  
J.~Phys.~A {\bf 20},  L97 (1987).

\bibitem{alc88} F.~C.~Alcaraz, M.~N.~Barber, and M.~T.~Batchelor, 
Ann. Phys. {\bf 182}, 280 (1988).

\bibitem{byers61} N. Byers and C.~N.~Yang, Phys. Rev. Lett. {\bf 7} 46 (1961).

\bibitem{shas90} B.~S.~Shastry, and B.~Sutherland, 
Phys. Rev. Lett. {\bf 65}, 243 (1990).

\bibitem{suth90} B.~Sutherland and B.~S.~Shastry, 
Phys. Rev. Lett. {\bf 65}, 1833 (1990).

\bibitem{ian01} I. Affleck and P. Simon, Phys. Rev. Lett. {\bf 86}, 2854 (2001).

\bibitem{wallin} M. Wallin, E.~S.~S\o rensen, S.~M.~Girvin and A.~P.~Young,
Phys. Rev. B {\bf 49}, 12~115 (1994).

\bibitem{privman}
\newblock V.~Privman, in 
\newblock {\em Finite Size Scaling and Numerical Simulation of
Statistical Systems}, ed.\ V.~Privman 
(World Scientific, Singapore, 1990), p.~1.

\bibitem{cardy84} J.~L.~Cardy, J. Phys. A {\bf 17}, L385 (1984); J.~L.~Cardy, Nucl. Phys. B {\bf 270}, 186 (1986).

\bibitem{affleck89} I.~Affleck, D.~Gepner, and H.~J.~Schulz,
J. Phys. A {\bf 22}, 511 (1989).

\bibitem{Schulz98} H.~J.~Schulz, G.~Cuniberti, and P.~Pieri,
''Fermi liquids and Luttinger liquids'', 
in {\it Field Theories for Low-Dimensional Condensed Matter Systems}, 
(Springer 2000).

\bibitem{voit95} J.~Voit, Rep. Prog. Phys.  {\bf 58}, 977 (1995).

\bibitem{KT73} J.~M.~Kosterlitz, and D.~Thouless, J. Phys. C {\bf 6}, 1181
(1973).

\bibitem{Affleck94} S.~Eggert, I.~Affleck, and M.~Takahashi,
Phys. Rev. Lett. {\bf 73}, 332 (1994).

\bibitem{cardy86}J.~L.~Cardy, J. Phys. A {\bf 19}, L1093 (1986).

\bibitem{Lukya98}S.~Lukyanov, Nucl. Phys. B {\bf 522}, 533 (1998).

\bibitem{Nomura93} K.~Nomura, Phys. Rev. B {\bf 48}, 16814 (1993).

\bibitem{Loss94} D.~Loss, and D.~L.~Maslov, 
Phys. Rev. B {\bf 74}, 178 (1995). 

\bibitem{baxter82} R.~J.~Baxter, 
{\it Exactly Solved models in Statistical Mechanics}, 
155 (Academic Press, 1982).


\bibitem{zotos}X.~Zotos, Phys. Rev. Lett. {\bf 82}, 1764 (1999).

\bibitem{alvarez}J.~V.~Alvarez and Claudius Gros, cond-mat/0105585

\bibitem{Affleck90}For a review and earlier references see I. Affleck, 
in Fields, Strings and Critical Phenomena, edited by E. Br\'ezin and J.  Zinn-Justin
(North Holland, Amsterdam, 1990), p. 563. 

\end{thebibliography}
\end{document}